\documentclass[12pt]{article}

\usepackage{latexsym,amsmath,amssymb,amsthm,amsfonts,graphicx,natbib}
\usepackage{epsfig}
\usepackage{bm}
\usepackage[pagewise,mathlines]{lineno}
\include{def}
\def\dispmuskip{\thinmuskip= 3mu plus 0mu minus 2mu \medmuskip=  4mu plus 2mu minus 2mu \thickmuskip=5mu plus 5mu minus 2mu}
\def\textmuskip{\thinmuskip= 0mu                    \medmuskip=  1mu plus 1mu minus 1mu \thickmuskip=2mu plus 3mu minus 1mu}
\def\beq{\dispmuskip\begin{equation}}    \def\eeq{\end{equation}\textmuskip}
\def\beqn{\dispmuskip\begin{displaymath}}\def\eeqn{\end{displaymath}\textmuskip}
\def\bea{\dispmuskip\begin{eqnarray}}    \def\eea{\end{eqnarray}\textmuskip}
\def\bean{\dispmuskip\begin{eqnarray*}}  \def\eean{\end{eqnarray*}\textmuskip}

\def\paradot#1{\vspace{1.3ex plus 0.7ex minus 0.5ex}\noindent{\bf\boldmath{#1.}}}

\newtheorem{theorem}{Theorem}

\newtheorem{proposition}[theorem]{Proposition}

\newcommand{\cdf}{cumulative distribution function}
\newcommand{\pdf}{probability density function}

\newcommand{\diag}{\text{diag}}

\newcommand{\half}{\frac{1}{2}}
\def\v{\boldsymbol}

\def\P{{\rm P}}                         % Probability

\def\v{\boldsymbol}

\def\G{\Gamma}
\def\a{\alpha}
\def\D{{\cal D}}
\def\d{\delta}
\def\g{\gamma}
\def\s{\sigma}
\def\k{\kappa}
\def\Sig{\Sigma}
\def\t{\theta}

\def\l{\lambda}
\def\Ld{\Lambda}

\def\LPDS{\text{\rm LPDS}}

\def\tr{\text{\rm tr}}

\def\diag{\text{\rm diag}}

\begin{document}
%\pagewiselinenumbers
\title{Copula-type Estimators for Flexible Multivariate Density Modeling using Mixtures}
\author{Minh-Ngoc Tran, Paolo Giordani, Xiuyan Mun, Robert Kohn, Mike Pitt
\footnote{M.-N. Tran (\texttt{minh-ngoc.tran@unsw.edu.au}), X. Mun (\texttt{z.mun@unsw.edu.au})
R. Kohn (\texttt{r.kohn@unsw.edu.au}) are at Australian School of Business,
University of New South Wales, Australia.
P. Giordani (\texttt{paolo.giordani@riksbank.se}) is at the Research Division, Swedish Central Bank, Sweden.
M. Pitt (\texttt{m.pitt@warwick.ac.uk}) is at the Economics Department, University of Warwick, UK.}}
\date{}
\maketitle

%======================================================================%
\begin{abstract}

Copulas are popular as models for multivariate dependence because they allow the marginal densities
and the joint dependence to be modeled separately. However, they usually require that the transformation from
uniform marginals to the marginals of the joint dependence structure is known. This can only be done
for a restricted set of copulas, e.g. a normal copula.
Our article introduces copula-type estimators for flexible multivariate density estimation which also
allow the marginal densities to be modeled separately from the joint dependence, as in copula modeling,
but overcomes the lack of flexibility of most popular copula estimators.
An iterative scheme is proposed for estimating copula-type estimators
and its usefulness is demonstrated through simulation and real examples.
The joint dependence is is modeled by mixture of normals  and mixture of normals factor analyzers models, and
mixture of $t$ and mixture of $t$ factor analyzers models. We develop efficient Variational Bayes
algorithms for fitting these in which model selection is performed automatically. Based on these
mixture models, we construct four classes of copula-type densities which are far
more flexible than current popular copula densities, and outperform them
in simulation and several real data sets.

\paradot{Keywords}
Mixtures of factor analyzers; Mixtures of normals;
Mixtures of $t$; Mixtures of $t$-factor analyzers; Variational Bayes.
\end{abstract}

\newpage

%======================================================================%
\section{Introduction} \label{Sec:introduction}
%======================================================================%
Multivariate density estimation is a fundamental problem in statistics and related fields.
One of the common approaches to multivariate density estimation is mixture modeling, which estimates the multivariate density of interest by
a multivariate mixture of densities such as a multivariate mixture of normal densities or a multivariate mixture of $t$ densities
\citep{Titterington:1985,McLachlan:2000}. Mixture models provide an automatic
method for estimating the density of non-standard and high-dimensional data.
 In principle, with sufficient data relative to the dimension
of the multivariate data, a mixture model can fit a data set arbitrarily well and capture most of its features.
In practice, however, transforming the marginals can greatly facilitate obtaining statistically efficient
estimates of a target multivariate density. This can be done informally by taking known transformations of the
marginals, for example by taking logs, or more formally, as we have done, by estimating the marginals flexibly and then
transforming.

A drawback in using mixture models is that we do not have much flexibility in modeling the
marginals, because all of the implied marginals are restricted to some particular form. For
example, if the multivariate density of interest is estimated by a multivariate mixture of normals then the
marginals of the target are estimated by the implied univariate mixture of normals. These
implied marginals may not even be close to the best models for the target marginals, which can be a
kernel density, a univariate mixture of $t$ or some parametric form. Furthermore, \cite{Giordani:2012}
observe that implicit estimation of marginals is in some cases less efficient
than direct estimation, even when the true model is used to fit the joint distribution. They
conjecture that the large number of parameters in the joint model that need to be estimated
makes the estimation practically less efficient, while direct estimation of the marginals does
not deteriorate with the dimension.

Copula modeling is a widely used approach to multivariate density estimation \citep{joe1997,Nelsen:1999}.
This approach is flexible in the sense that it allows one to model
the marginals and the joint dependence separately.
Because of computational reasons,
the joint dependence is often estimated by a mathematically convenient model
such as a multivariate normal or a multivariate $t$ distribution.
Such conveniently parametric copula models may not be appropriate
for modeling data sets that have a complex joint dependence structure.
For example, different areas in the domain of the data may have different dependence structures
(see the motivating example in Section \ref{Sec:generalized copula} and the Iris data in Section \ref{Sec:applications}).
In such cases, a multivariate mixture model will capture the joint dependence of the data better than a simple model such as a normal or a $t$ model.
It is therefore desirable to use flexible models such as multivariate mixture models
to estimate the joint dependence.

This article proposes a new class of multivariate density estimators called copula-type estimators
which have the motivation
of using flexible models for estimating complex joint dependence structure,
while preserving the possibility offered by copulas of modeling the marginal distributions separately.
Except in some special cases, copula-type estimators are not copula estimators,
although they still allow the marginals to be separately estimated.
The construction of copula-type estimators allows us to estimate them using an iterative scheme.
The construction also covers many popular copula estimators found in the literature.
The article focuses on a class of copula-type estimators
using multivariate mixture models to capture the joint dependence of the target density.
In particular, four copula-type estimators are considered: a copula-type estimator based on a multivariate mixture of normals,
a copula-type estimator based on a multivariate mixture of $t$,
a copula-type estimator based on a mixture of factor analyzers
and a copula-type estimator based on a mixture of $t$-factor analyzers.
These four copula-type estimators allow us to achieve flexibility, efficiency and robustness in multivariate density estimation.
Their  estimation is based on efficient Variational Bayes algorithms
for fitting mixture models, in which model selection (and factor selection) is automatically incorporated.
See, e.g., \cite{Ormerod:2009} for an introduction to the Variational Bayes method.
We  believe that our algorithm for fitting mixtures of mixtures of $t$ and $t$-factor analyzers is the first method in the literature
which is able to do parameter estimation and component and factor selection simultaneously and automatically.

The article is organized as follows.
Section \ref{Sec:generalized copula} presents the main results.
%contains the following:
%(i) a brief review of copula modeling and a motivating example demonstrating the disadvantages of parametric copulas;
%(ii) a description of our framework for constructing copula-type estimators;
%(iii) some properties of copula-type estimators;
%(iv) an iterative scheme for estimating copula-type estimators;
%and (v) four particular copula-type estimators using multivariate mixtures.
%Section \ref{Sec:mixture} presents mixture models and our Variational Bayes fitting method.
%%Section \ref{Sec:MA} presents the marginal adaptation approach and introduces two new marginally adjusted estimators.
%Section \ref{Sec:marginals} discusses marginal estimation.
Section \ref{Sec:applications} presents a simulation study
and several applications to real data.
Section \ref{Sec:conclusion} concludes the article.
Proofs and technical details are presented in the Appendices.

%======================================================================%
\section{The copula-type model} \label{Sec:generalized copula}
%======================================================================%
%------------------------------------------------%
\subsection{Copula modeling}
%------------------------------------------------%
Suppose that we are given a data set
$\mathcal D_Y=\{\v y_i=(y_{i1},...,y_{id})',\ i=1,...,n\}$
of realizations of a random vector $\v Y=(Y_1,...,Y_d)'$,
and we wish to estimate the distribution of $\v Y$.
We will denote random variables by upper-case letters,
their realizations by lower-case letters,
and write vector variables in bold.
We write $\v y$ for a general multivariate argument
and $\v y_i$ for a particular realization.
We restrict the discussion in this paper to continuous marginals.

In copula modeling, one often assumes that $\v Y=(Y_1,...,Y_d)'$
inherits the joint dependence structure from another continuous random vector $\v X=(X_1,...,X_d)'$.
Let $G(\v x)$ be the joint \cdf{} (cdf) of $\v X$ and $G_j(x_j)$, $j=1,...,d$, be its marginal cdf's.
Write the corresponding \pdf{s} (pdf's) as $g(\v x)$ and $g_j(x_j), j=1, \dots, d$.
The joint dependence of $\v Y$ is assumed to be constructed from $\v X$ as follows.
First, let $U_j=G_j(X_j)$, $j=1,...,d$.
Each $U_j$ has a uniform distribution on $[0,1]$
while their joint dependence is induced from that of $G$,
i.e. the cdf of $\v U$ can be written as
\beq\label{copula}
C(\v u|G) = G(G_1^{-1}(u_1),...,G_d^{-1}(u_d)),\;\;\v u=(u_1,...,u_d)'.
\eeq
This function is referred to as a copula function or a copula (induced by $G$).
This way of constructing a copula is known as the inverse method \citep{Nelsen:1999}.

Given univariate (continuous) cdf's $F_1,...,F_d$, let $Y_j=F_j^{-1}(U_j)$, $j=1,...,d$.
Then each random variable $Y_j$ admits $F_j$ as its cdf while their joint dependence
is induced from that of the vector $\v X$,
i.e. the cdf $F$ of $\v Y$ can be expressed in terms of $G$ as
\beq\label{Ycdf}
F(\v y) = C(F_1(y_1),...,F_d(y_d)|G) = G\Big(G_1^{-1}(F_1(y_1)),...,G_d^{-1}(F_d(y_d))\Big).
\eeq
We refer to $F(\v y)$ (or its pdf $f(\v y)$) as a copula cdf, which can be though of as an
approximation to the true cdf of $\v Y$.
It is easy to see that the $i$th marginal cdf of $F$ is $F_i$.
Figure \ref{F:diagram} demonstrates this $\v X\leftrightarrow\v U\leftrightarrow\v Y$
and $G\leftrightarrow C\leftrightarrow F$ relationship diagrammatically.
The three random vectors $\v X$, $\v U$ and $\v Y$ have different marginals but
the same joint dependence structure in the sense that their cdf's can be written in terms of the
copula $C$.

\begin{figure}[h]
\centering
\setlength{\unitlength}{1mm}
\begin{picture}(50,45)
\put(0,40){$\v X=(X_1,...,X_d)'\sim G(\v x),\ X_j\sim G_j(x_j)$}
\put(30,37){\vector(0,-1){12}}
\put(30,25){\vector(0,1){12}}
\put(32,30){$U_j=G_j(X_j)$}
\put(0,20){$\v U=(U_1,...,U_d)'\sim C(\v u|G),\ U_j\sim U[0,1]$}
\put(30,17){\vector(0,-1){12}}
\put(30,5){\vector(0,1){12}}
\put(32,10){$Y_j=F_j^{-1}(U_i)$}
\put(0,0){$\v Y=(Y_1,...,Y_d)'\sim F(\v y),\ Y_j\sim F_j(y_j)$}
\end{picture}
\caption{$\v X\leftrightarrow\v U\leftrightarrow\v Y$ and $G\leftrightarrow C\leftrightarrow F$ relationship. $\v X$, $\v U$ and $\v Y$
have the same joint dependence structure but different marginals.
}
\label{F:diagram}
\end{figure}

Two examples of popular copulas are the normal and $t$ copulas.
In the normal copula $G$ is assumed to be the cdf of a multivariate normal distribution $N_d(\v 0,V)$,  with
 $G$ is assumed to be the cdf of a multivariate $t$ distribution $t_d(\v0,\nu,V)$ with $\nu$ the degrees of freedom and
$V$ is a scale matrix with diagonal entries 1. For both the normal and $t$ copulas the scale matrix $V$ is a correlation matrix.

Inference in copula modeling consists of two problems.
The first is how to estimate the marginal cdf's $F_j$
and the second is how to select and estimate an appropriate copula $C$, or equivalently $G$.
This section focuses on the second problem, i.e. on estimating an appropriate joint dependence structure.
We assume for now that the marginal cdf's $F_j$ are known;
marginal estimation is discussed in Section \ref{Sec:marginals}.
By making the transformation $u_{ij}=F_j(y_{ij})$, $j=1,...,d$, $i=1,...,n$,
we obtain a data set $\mathcal D_U=\{\v u_i, i=1,...,n\}$ in the $\v U$-space
and the problem reduces to reconstructing the source of dependence structure in $\v X$ based on $\mathcal D_U$.
It is worth emphasizing that the data $\D_U$ contain all information we have about the joint dependence of $F$ (or $G$).

The main problem with many current approaches for fitting joint dependence using
copulas is that if an inappropriate choice of copula is made, then the transformed data in the $\v X$-space
may be harder to model than the original data $\D_Y$. The following discussion and example consider this issue.
Suppose that we wish  to estimate the joint dependence in $\D_U$
by a multivariate cdf $\widehat G$, where $\widehat G$ is assumed known up to some parameters
that need to be estimated from the data.
For example, $\widehat G$ may be a multivariate normal cdf whose mean is $\v 0$ and whose covariance
matrix is a correlation matrix that needs to be estimated from the data.
We further assume that the marginal cdf's $\widehat G_j$ of $\widehat G$ are fully known.
This is the case, for example, in the normal copula or the $t$ copula with fixed degrees of freedom.
Then a simple method for estimating $\widehat G$ is as follows:
first, transform the data $\D_U$ to a data set $\D_X^{\widehat G_{1:d}}$ in the $\v X$-space via $x_{ij}=\widehat G_j^{-1}(u_{ij})$, $j=1,..,d$, $i=1,...,n$;
then, fit $\widehat G$ to $\D_X^{\widehat G_{1:d}}$.
For example, in fitting a normal copula we first make the transformation $x_{ij}=\Phi^{-1}(u_{ij})$
with $\Phi$ the standard normal cdf and then fit a multivariate normal distribution $N_d(\v0,V)$ (with $V$ a correlation matrix)
to this $\v X$-space data set.
The idea (hope) is that the transformed data $\D_X^{\widehat G_{1:d}}$ are easier to model than $\D_Y$.
However, in some cases $\D_X^{\widehat G_{1:d}}$ cannot be fitted well by $\widehat G$.
The main problem with copulas is that
with an inappropriate choice of $\widehat G$,
the transformed data may be harder to model
than the original data $\D_Y$.
This is illustrated in the example below.

\paradot{A motivating example}
We construct a two-dimensional vector $\v Y$ whose joint dependence is induced from another vector $\v X$ as in Figure \ref{F:diagram}.
$\v X$ is distributed as a multivariate mixture of two normals with density
\beq\label{equ:DGP}
g(\v x) = 0.5N_2(\v\mu_1,V_1)+0.5N_2(\v\mu_2,V_2),
\eeq
where
\beqn
\v\mu_1=\begin{pmatrix}2\\2 \end{pmatrix},\;\;
\v\mu_2=\begin{pmatrix}-2\\-2 \end{pmatrix},\;\;
V_1=\begin{pmatrix}1&0.6\\0.6&1 \end{pmatrix},\;\;
V_2=\begin{pmatrix}1&-0.6\\-0.6&1 \end{pmatrix},
\eeqn
and $Y_1\sim N_1(1,3)$ and $Y_2\sim t_1(0,1,5)$.

\begin{figure}[h]
\centering
\includegraphics[width=1\textwidth,height =.5\textheight]{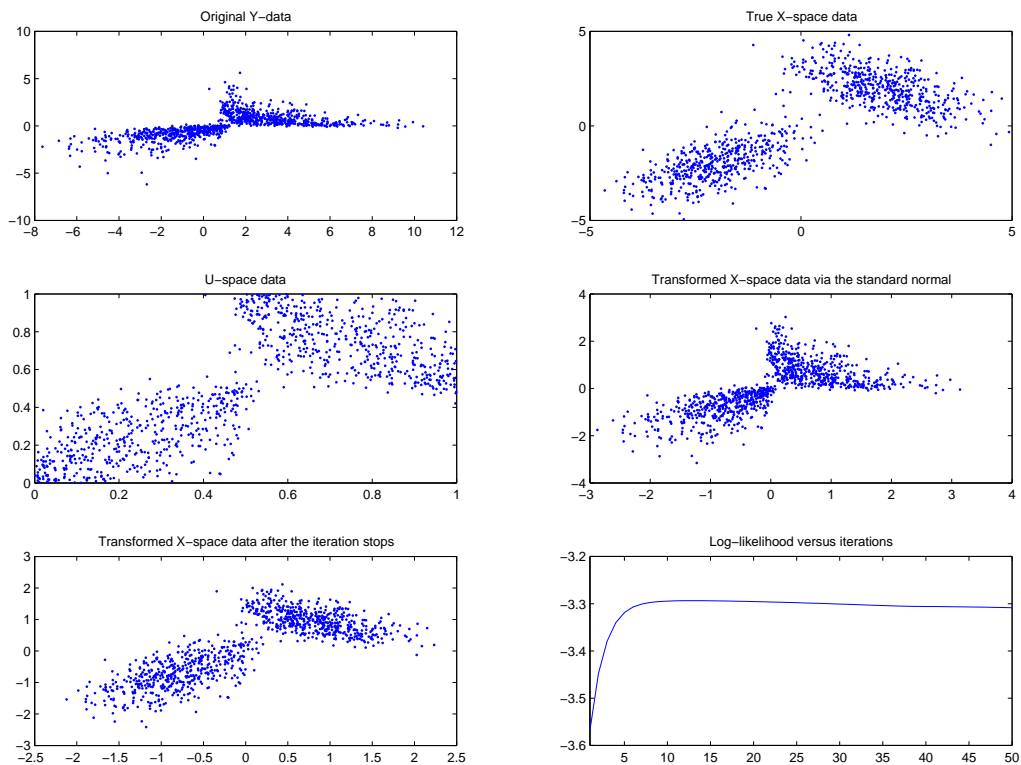}
\caption{Motivating example: The first row shows the original $\v Y$-data and the true dependence structure in the $\v X$-space,
which is equivalently transformed to the $\v U$-space (middle-left panel) via $u_{ij}=G_j(x_{ij})$.
The middle-right panel shows the transformed data $\D_X^{\widehat G_{1:d}}$ when $\widehat G$ is a normal distribution.
The bottom-left panel shows the transformed data $\D_X^{H_{1:d}}$ when the $H_j$ are mixtures of two normals obtained by the iterative scheme.
The last panel plots the log-likelihood values versus iterations.
}
\label{F:Figure1}
\end{figure}

The panels in the first row of Figure \ref{F:Figure1} show 1000 realizations from the vectors $\v Y$ and $\v X$ respectively.
The left panel of the middle row plots the data $\D_U$ obtained via $u_{ij}=G_j(x_{ij})$,
which contain all information about the joint dependence of $\v Y$ (and $\v X$).
If we use a normal copula to model the dependence structure in $\v Y$,
we need to fit a bivariate normal distribution to the data shown in the right panel of the middle row,
which are obtained via $x_{ij}=\Phi^{-1}(u_{ij})$.
Clearly a multivariate normal density does not provide a good fit to this data set
and it is necessary to have a more flexible model than a multivariate normal distribution to capture the joint dependence.

The example above motivates the use of flexible models to estimate the joint dependence.
Suppose that $G(\v x)=G(\v x|\v\t)$ belongs to some class of multivariate cdf's,
such as the cdf's of multivariate mixtures of normals, with unknown parameter vector $\v\t$.
From \eqref{Ycdf}, the pdf of $\v Y$ is
\beq\label{Ypdf}
f(\v y|\v\t)=\frac{g(\v x|\v\t)}{\prod_{j=1}^dg_j(x_j|\v\t)}\prod_{j=1}^df_j(y_j),
\eeq
where $x_j=G_j^{-1}(F_j(y_j)|\v\t)$.
It is possible, in principle,  to estimate $\v\t$ by maximum likelihood or by its posterior mode based on the pdf \eqref{Ypdf}.
However, when $G(\v x|\v\t)$ is a complex cdf such as a mixture cdf,
optimization over $\v\t$ is computationally very difficult, for two reasons.
First, we  cannot in general compute the gradient of the likelihood analytically because the pdf \eqref{Ypdf} has $\v\t$ deeply embedded
in the inverse transformations $x_j=G_j^{-1}(F_j(y_j)|\v\t)$.
Second, this is a high-dimensional optimization problem with the complex constraints that the scale correlation matrices in the mixture
need to be positive definite.
For example, suppose that $G(\cdot|\v\t)$ is the cdf of a mixture of $K$ normals; then the dimension of $\v\t$ is $\text{dim}= K-1
+ dK + \half d(d+1)K = \half K (d+1)(d+2)-1$,
which can be thousands for even a moderate $d$; here, we have $K$ components, $K-1$ probability parameters,  $Kd$ mean parameters and
$Kd(d+1)/2 $ correlation parameters. We note that we tried black box optimization in Matlab for a two dimensional ($d=2$)
problems, but the optimization algorithm repeatedly failed to converge.

In the next section we propose a class of copula-type (CT) estimators which estimate the marginals from $\D_Y$ as well as flexibly estimating
the dependence structure.

%------------------------------------------------%
\subsection{Copula-type estimators}
%------------------------------------------------%
We now describe a framework for constructing flexible multivariate density estimators,
which allows using complex and flexible models for estimating the joint dependence structure.
Note that we are assuming that the marginal cdf's $F_j(y_j)$ are given or separately estimated,
so that we start with the transformed data $\D_U$ and wish to capture the {\em joint dependence} of $\v X$.

Our estimator for the distribution of interest is constructed as follows.
Suppose that univariate cdf's $H_j$ are an initial guess of the marginal cdf's $G_j$, $j=1,...,d$.
Recall that $G$ is the cdf of $\v X$ and $G_j$ are its marginal cdf's,
$G$ is unknown and we wish to estimate $G$.
Let $\D_X^{H_{1:d}}$ be the data set in the $\v X$-space obtained by transforming $x_{ij}=H_j^{-1}(u_{ij})$.
Now fit a multivariate cdf $\widehat G$ to $\D_X^{H_{1:d}}$.
For example, $H_j$ can be the cdf of an univariate mixture of normals
and $\widehat G$ the cdf of a multivariate mixture of normals.
Let
\beq\label{e:GC}
\widehat C(\v u|H,\widehat G)=\widehat G(H_1^{-1}(u_1),...,H_d^{-1}(u_d)).
\eeq
We note that $\widehat G $ is selected from the class of cdf's corresponding to mixture of normals, mixture of factor
analyzers, mixtures of $t$ and mixture of $t$ analyzers. That is, $\widehat G $ is specified up to class, e.g. mixture of normals,
with the parameters, number of components and number of factors unknown and to be estimated form the data.

The following result provides an explicit expression for the estimator.
\begin{proposition} \label{P: propotion dist y}
The cdf of the estimator for the distribution of $\v Y$ is
\beq\label{e:GCcdf}
\widehat F(\v y|H,\widehat G) = \widehat C(F_1(y_1),...,F_d(y_d)|H,\widehat G).
\eeq
The pdf of the estimator is
\beq\label{GCestimator}
\widehat f(\v y|H,\widehat G)=\widehat g(\v x)\prod_{j=1}^d\frac{f_j(y_j)}{h_j(x_j)},
\eeq
its $j$th marginal pdf is
\beqn
\widehat f_j( y_j|H,\widehat G)=\frac{\widehat g_j(x_j)}{h_j(x_j)}f_j(y_j)
\eeqn
with $x_j=H_j^{-1}(F_j(y_j))$ and $\widehat g$, $\widehat g_j$, $f_j$, $h_j$ density functions
with respect to $\widehat G$, $\widehat G_j$, $F_j$, $H_j$ respectively.
\end{proposition}

We note that equation \eqref{e:GC} is not necessarily a copula.
It is also important to note that $\widehat f(\v y|H,\widehat G)$ in \eqref{GCestimator}
is a valid multivariate density for any $\widehat g$, $h_j$ and $f_j$.
To see this, using the equality that $h_j(x_j)dx_j=f_j(y_j)dy_j$,
we can prove that $\int\widehat f(\v y|H,\widehat G)d\v y=1$.
This justifies the stopping criterion used in the iterative scheme in Section \ref{Subs:iterative}.

The following result guarantees that under some conditions the marginals of the estimator $\widehat f$
converge to the true marginals $f_j$.
We say that a fitting method is reliable if the resulting estimator $\widehat g(\v x)$ converges in total variation norm
to the underlying density $h(\v x)$
% h^{H_{1:d}} (\v x)$
that generates the data ${\cal D}_X^{H_{1:d}} $ , i.e.
\beqn
d_\text{TV}(\widehat g,h)=\frac12\int |\widehat g(\v x)-h(\v x)|d\v x\to 0,
\eeqn
as the sample size increases.
\begin{proposition}[Marginal consistency]\label{P:marginal}
Suppose that the method for fitting $\widehat G$ to $\D_X^{H_{1:d}}$ is reliable.
%in the sense that
%the estimator $\widehat g(\v x)$ tends to
%the density that generates $\D_X^{H_{1:d}}$ with probability 1 as the sample size increases.
Then $\widehat f_j( y_j|H,\widehat G)$ converges in total variation to the true marginal $f_j(y_j)$, $j=1,..,d$,
as the sample size increases.
\end{proposition}
The proofs of the two  propositions are in Appendix A.
%Note that the assumption in Proposition \ref{P:marginal} does not require
%$\widehat g(\v x)$ to converge to $g(\v x)$, which would be a very strong assumption.
%It only requires that $\widehat g(\v x)$ converges to the density that governs $\D_X^{H_{1:d}}$,
%which is a  milder assumption if we have a rich enough model for $\D_X^{H_{1:d}}$.
%That is, in large samples, the fitting method is able to produce density estimators that converge to the true density.

We call the function $\widehat C$ in \eqref{e:GC} a copula-type function,
and refer to \eqref{e:GCcdf} or \eqref{GCestimator} as a copula-type estimator.
This is because $\widehat C$ has a similar form as the copula function $C$ in \eqref{copula},
and under some conditions (see below) a copula-type function becomes a copula function.

This approach to multivariate density estimation is flexible for the following reasons.
\begin{itemize}
\item It allows us to use complex and principled models such as multivariate mixture models to
estimate the joint dependence.
\item With appropriate choices of the $H_j$ and $\widehat G$, the framework covers some popular copulas in the literature.
For example, with $H_j=\Phi$, $\widehat G=N_d(\v 0,V)$ and $V$ a correlation matrix we obtain the normal copula model;
with $H_j=t_1(0,\nu,1)$, $\widehat G=t_d(\v0,\nu,V)$ and $V$ a scale matrix with diagonal entries 1 we obtain the $t$ copula model.
Note that in these two cases, $\widehat f_j\equiv f_j$, $j=1,...,d$.
More generally, a copula-type function is a copula function if $\widehat G$ admits $H_j$'s as its marginal cdf's.
\item If $H_j\equiv F_j$, then $\widehat G$ is fit directly to the original data, i.e. no marginal adaptation is used.
Then copula-type modeling reduces to the usual multivariate modeling, such as multivariate mixture modeling.
\end{itemize}
We note that unless $\widehat G_j=H_j$, copula-type estimators are not true copula estimators
because the marginal pdf's $\widehat f_j( y_j|H,\widehat G)$ of a copula-type estimator
are not exactly the separately estimated marginal pdf's $f_j$.
In order for a copula-type estimator to be a copula estimator
it is necessary to impose the constraint $\widehat G_j=H_j$.
However, imposing this constraint usually makes the estimation of $\widehat G$ very difficult,
especially when complex models are used to estimate the joint dependence.
Furthermore, this constraint need not lead to better performance; see the remarks at the end of Section \ref{Subs:iterative}.
Finally, Proposition~\ref{P:marginal} guarantees that in large samples a copula-type estimator converges to an exact copula estimator
if the model for $\D_X^{H_{1:d}}$ is sufficiently flexible.

%------------------------------------------------%
\subsection{Iterative scheme}\label{Subs:iterative}
%------------------------------------------------%
In general, we should choose the univariate cdf $H_j$ such that the transformed data $\D_X^{H_{1:d}}$
look as if they can be effectively fitted by the candidate set of multivariate distributions $\widehat G$.
In our case, this means that the
transformed data can be parsimoniously fitted by a multivariate mixture of
normals or a multivariate mixture of $t$.
This is difficult if the $H_j$ are only chosen once.
We propose an iterative scheme which is useful for estimating the $H_j$ and $\widehat G$ in general.
Assume that $\widehat G(\v x)$ belongs to some family of multivariate cdf's
such as multivariate normal mixture cdf's: $\widehat G(\v x)=\widehat G(\v x|\v\t)$ with $\v\t$ the parameters.
We start with some initial univariate cdf's $H_j(x_j)=H_j^{(0)}(x_j)$,
fit $\widehat G(\v x|\v\t)$ to $\D_X^{H_{1:d}}$ to get an estimate $\widehat{\v\t}$ of $\v\t$ and then repeat the procedure
with $H_j(x_j)$ set to $\widehat G_j(x_j|\widehat{\v\t})$.
\begin{enumerate}
\item Start with some initial univariate cdf's $H_j(x_j)=H_j^{(0)}(x_j)$.
\item Transform the data $\D_U$ to $\D_X^{H_{1:d}}$ via $x_{ij}=H_j^{-1}(u_{ij})$, $j=1,...,d$, $i=1,...,n$.
\item Fit $\widehat G(\v x|\v\t)$ to $\D_X^{H_{1:d}}$ to get an estimate $\widehat{\v\t}$ of $\v\t$.
\item Set $H_j(x_j)=\widehat G_j(x_j|\widehat{\v\t})$ with $\widehat G_j(x_j|\widehat{\v\t})$ the $j$th marginal cdf of $\widehat G(\v x|\widehat{\v\t})$. Go back to Step 2.
\end{enumerate}
We suggest stopping the iteration if the log-likelihood
\beqn
\sum_{\v y\in\D_Y}\log\widehat f(\v y|H,\widehat G)
\eeqn
does not improve any further. The iteration uses Variational Bayes at each iteration to choose
the parameters, as well as choosing automatically the number of components and number of factors.
Note that $\widehat f(\v y|H,\widehat G)$ is a valid density.
We observe that the log-likelihood often increases in the first few iterations and then decreases;
see the last panel in Figure \ref{F:Figure1}.
A possible choice for the initial marginal distributions $H_j^{(0)}$ is the standard normal cdf $\Phi$.
When $\widehat G$ is the cdf of a multivariate mixture of normals
or a multivariate mixture of $t$,
we suggest selecting the $H_j^{(0)}$ as the implied marginals of the multivariate mixture distribution
estimated from the original data $\D_Y$.
We found that the resulting estimates are insensitive to the initial distribution taken and
show the usefulness of this scheme through numerical examples.

\paradot{A motivating example (continued)}
We now apply the iterative scheme to estimate the joint dependence in $\v Y$
with $H_j^{(0)}(x_j)=\Phi(x_j)$ and $\widehat G(\v x|{\v\t})$ a multivariate mixture of two normals.
The procedure stops after 13 iterations when the log-likelihood is maximized.
The bottom right panel in Figure~\ref{F:Figure1} plots the log-likelihood values vs  iterations number.
The bottom left panel shows the transformed data $\D_X^{H_{1:d}}$ after the iterative scheme stops.
Clearly, the {\em joint dependence structure} of this estimated $\widehat G$ is similar to that of the true distribution $G$.
In fact, the two component correlation matrices of $\widehat G$ are $[1\ 0.62;\ 0.62\ 1]$ and $[1\ -0.59;\ -0.59\ 1]$,
which are close to the true matrices $V_1$ and $V_2$.

\paradot{Remark 1}
A different, but related, estimator constructed
within our framework is
\beq\label{altGCestimator}
\widehat f(\v y|\widehat G)=\widehat g(\v x)\prod_{i=1}^d\frac{f_j(y_j)}{\widehat g_j(x_j)},
\eeq
with $\widehat G$ obtained after the iteration above has terminated,
i.e. we use the copula induced by $\widehat G$ to construct the estimator.
The estimator \eqref{altGCestimator} is a copula estimator as its marginals are equal to $f_j$.
However, our experiments show that the copula-type estimator \eqref{GCestimator}
usually has a slightly better performance in terms of
the log predictive density score (see Section \ref{Sec:applications}) than the copula estimator \eqref{altGCestimator}.
We conjecture that this is because the expression \eqref{GCestimator} takes into account
the actual marginal transformations of the data $H_j(x_j)=F_j(y_j)$,
while \eqref{altGCestimator} uses only the estimated joint dependence.

\paradot{Remark 2} The proposed method can be easily extended to the case
where the marginals depend on covariates $\v z$.
Assume that $\{(\v y_i,\v z_i),\ i=1,...,n\}$ are observations
from a multivariate distribution $F(\v y|\v z)$,
whose joint dependence is independent of $\v z$.
Let $u_{ij}=F_j(y_{ij}|\v z_i)$, $j=1,...,d$, $i=1,...,n$,
where $F_j(y_j|\v z)$ is the $j$th marginal cdf.
We can now use the iterative scheme to estimate the $H_j$ and $\widehat G$.
The pdf of the estimator is expressed as
\beqn
\widehat f(\v y|\v z, H,\widehat G)=\widehat g(\v x)\prod_{j=1}^d\frac{f_j(y_j|\v z)}{h_j(x_j)},
\eeqn
with $x_j=H_j^{-1}(F_j(y_j|\v z))$.
Extension to the case where the distribution functions $C$ and $G$ depend on covariates
is more difficult and is left for future research.

%-------------------------------------------%
\subsection{Copula-type estimators based on mixtures}\label{Sec:FC-mixtures}
%-------------------------------------------%
This paper considers in particular four copula-type distributions using
multivariate mixture models to estimate the joint dependence.

The first copula-type  estimator uses  a multivariate mixture of normals to model the joint dependence
and is denoted by CT-MN.
See, e.g., \cite{Titterington:1985} and \cite{McLachlan:2000} for an introduction to mixture models.

A mixture of normals model may be over-parameterized when modeling high-dimensional data
as the number of model parameters increases at least quadratically with the dimension.
This is because the number of parameters in
each component increases quadratically and
the number of components is also likely to increase with dimension.
Parameter estimation is typically less efficient statistically if the number of observations is small relative to the number of parameters.
In such cases, it is desirable to reduce the number of parameters.
The mixture of factor analyzers model introduced in \cite{Ghahramani:1997}
provides an effective way to parsimoniously
model high-dimensional data,
and inherits the advantages of flexibility from mixture modeling
and dimensionality reduction from the factor representation.
The second copula-type estimator is based on a mixture of factor analyzers
and is denoted by CT-MFA.

\cite{krupskii:joe:2013} propose a general one component factor copula model which they propose
to estimate by maximum likelihood. However, they do not address the two main issues in our article, i.e.
the estimation of a copula of a mixture and how to make the marginals in that copula consistent
with the joint distribution.

The third copula-type estimator uses a multivariate mixture of $t$ to model the joint dependence and is
denoted as CT-M$t$.
The heavy tails of $t$ distributions can make this estimator successful when modeling data
with outliers or atypical observations.
See, e.g., \cite{Peel:2000} for a discussion of the multivariate mixture of $t$ model.
The fourth copula-type uses a mixture of $t$-factor analyzers to estimate the joint dependence,
and we denote it by CT-M$t$FA.

We note that our component and factor selection approach will indicate if a
 simpler normal or t copula or a factor version of these models is sufficient to fit the data. We can
also use cross-validation log predictive score (LPDS -- see the definition in Section \ref{Sec:applications})
to make a similar assessment.

Appendix B presents more
details on mixture modeling and how to fit a mixture model to the data
using Variational Bayes methods.

%======================================================================%
\subsection{Estimation of marginals}\label{Sec:marginals}
%======================================================================%
Estimating the marginal densities of $\v Y $ is typically much easier than estimating the joint dependence structure.
There are a number of efficient approaches for estimating a univariate density,
for example parametric estimation, kernel density estimation and univariate mixture estimation.
Given a class $\mathcal F$ of univariate density estimators, the best estimator can be selected using cross validation LPDS (see Section
 \ref{Sec:applications}).
In the examples below, we consider for the class $\mathcal F$
a kernel density estimator, a univariate mixture of normals estimator, a univariate mixture of $t$ estimator,
an implied univariate mixture of normals estimator (i.e. the univariate density estimator for the marginal implied from
the multivariate mixture of normals for the joint)
and an implied univariate mixture of $t$ estimator. When fitting univariate mixtures to the marginals,
the number of components is selected by Variational Bayes for the real examples.
For the simulated example, we used the true model that generated the data as the best model for the marginals because
we wished to focus on how well the joint density was being estimated.

%======================================================================%
\section{Examples} \label{Sec:applications}
%======================================================================%
A common measure for the performance of a density estimator
is the log predictive density score (LPDS) \citep[see, e.g.,][]{Good:1952,Geisser:1980}.
Let $\D_T$ be a test data set that is independent of the training set $\D$.
Suppose that $\widehat p(\v y|\D)$ is a density estimator based on $\D$.
The LPDS of the estimator $\widehat p$ is defined by
\beqn
\LPDS(\widehat p)=\frac{1}{|\D_T|}\sum_{\v y_i\in \D_T}-\log\widehat p(\v y_i|\D_T),
\eeqn
with $|\D_T|$ the number of observations in $\D_T$.
The smaller the LPDS the better the estimator.

For the real examples considered in this section
we use the cross-validated LPDS.
Suppose that the data set $\D$ is split into roughly $B$ equal parts $\D_1,...,\D_B$,
the $B$-fold cross-validated LPDS is defined as
\beqn
\LPDS(\widehat p)=\frac{1}{|\D|}\sum_{j=1}^B\sum_{\v y_i\in \D_j}-\log\widehat p(\v y_i|\D\setminus \D_j).
\eeqn
When computing this cross-validated LPDS, the marginal models are fixed at the best models which have been already selected (again, by cross-validated LPDS for each marginal). That is, the models for the marginals and the copula model for the joint are specified up to class with the parameters (including the number of components and number of factors) estimated from each data set $\D\setminus \D_j$.
We take $B = 5$ or $B = 10$ as recommended by \cite{hastie:etal:2009}, pp. 241-244.

\cite{Giordani:2012} propose a class of multivariate density estimators
to improve on standard multivariate estimators.
They do so by allowing the user to adjust any initial multivariate estimator by the best fitting density for each marginal.
\cite{Giordani:2012} introduce two marginally adjusted estimators using the mixture of normals and mixture of factor analyzers models for the initial estimators.
These estimators are denoted by MAMN and MAMFA.
A total of 12 estimators are considered below for comparison.
The first six are mixture-based estimators including
a multivariate mixture of normals (MN),
a multivariate mixture of $t$ (M$t$),
a mixture of factor analyzers (MFA),
a mixture of $t$-factor analyzers (M$t$FA),
and two marginally adjusted estimators, MAMN and MAMFA.
The others are copula-based estimators including
a normal copula (NC), a $t$ copula ($t$C), CT-MN, CT-M$t$, CT-MFA and CT-M$t$FA.

%---------------------------------------------%
\subsection{Simulated Example}
%---------------------------------------------%
We consider the data generating process as in the motivating example in Section \ref{Sec:generalized copula}.
Given a dimension $d$, a training data set $\D$ of size $n$ is generated from \eqref{equ:DGP},
where $\v\mu_1=(-2,...,-2)'$ and $\v\mu_2=(2,...,2)'$ are vectors of size $d$,
$V_1=(V_{1,ij})_{i,j}$, $V_2=(V_{2,ij})_{i,j}$ with $V_{1,ij}=0.5^{|i-j|}$ and $V_{2,ij}=(-0.5)^{|i-j|}$,
and $Y_j\sim t_1(0,1,5)$ for all $j=1,...,d$.
A test data set $\D_T$ of 1000 realizations is then generated in the same manner to compute the log predictive density scores.
For each  $d$ and $n$ combination, we compute the 12 density estimators based on $\D$, their LPDS based on $\D_T$, CPU times,
and replicate this computation for 50 replications.
Tables \ref{T:simulation_LPDS} and \ref{T:simulation_CPU} summarize the LPDS and CPU times averaged over the replications for various $d$ and $n$.

We draw the following conclusions.
1) The copula-type estimators perform best,
except for the CT-M$t$ when $d=40$ and $n=200,\ 500$.
We conjecture that estimating the CT-M$t$ in the large-$d$ small-$n$ case
is challenging because of a very large number of parameters that need to be estimated.
We observe that the CT-M$t$ works well when $n$ is large enough.
2) Dimension reduction via the factor analyzers models is useful when $d$ is large.
3) The marginally adjusted estimators, MAMN and MAMFA, always outperform their initial estimators, MN and MFA.
4) The normal and $t$ copulas work poorly.
This is not surprising as the joint dependence of the data has a mixture structure.
5) The copula-type estimators are more time consuming
than the others, principally because in the variational Bayes algorithms, components (and factor) selection
takes place every iteration. The code is written in Matlab and run on an Intel Core
16 i7 3.2GHz desktop.

\begin{table}[h]
\renewcommand{\arraystretch}{0.9} \centering \fontsize{7.5pt}{12pt}\selectfont
\centering
\vskip1mm
\begin{tabular}{cc|cccccccccccc}
$d$	&$n$	&MN	&M$t$	&MFA	&M$t$FA	&MAMN	&MAMFA	&NC	&$t$C	&CT-MN	&CT-M$t$	&CT-MFA	&CT-M$t$FA\\
\hline
5	&200	&5.01	&4.96	&5.17	&4.97	&4.94	&5.15	&5.62	&5.62	&4.45	&{\bf 4.33}	&4.66	&4.34\\
	&	&(0.07)	&(0.06)	&(0.06)	&(0.04)	&(0.08)	&(0.06)	&(0.04)	&(0.04)	&(0.09)	&(0.08)		&(0.09)	&(0.08)\\
	&500	&4.84	&4.80	&5.05	&4.85	&4.78	&5.03	&5.53	&5.53	&4.19	&{\bf 4.15}	&4.35	&4.19\\
	&	&(0.06)	&(0.06)	&(0.06)	&(0.06)	&(0.05)	&(0.05)	&(0.07)	&(0.07)	&(0.08)	&(0.06)		&(0.16)	&(0.07)\\
\hline
10	&200	&9.27	&9.31	&9.36	&9.09	&9.11	&9.26	&10.47	&10.47	&8.03	&8.09		&8.14	&{\bf 7.90}\\
	&	&(0.16)	&(0.12)	&(0.19)	&(0.13)	&(0.15)	&(0.24)	&(0.15)	&(0.15)	&(0.16)	&(0.19)		&(0.46)	&(0.18)\\
	&500	&8.85	&8.80	&9.03	&8.80	&8.73	&8.97	&10.22	&10.22	&7.74	&{\bf 7.44}		&7.65	&7.47\\
	&	&(0.11)	&(0.12)	&(0.13)	&(0.10)	&(0.11)	&(0.12)	&(0.04)	&(0.04)	&(0.20)	&(0.10)		&(0.19)	&(0.08)\\
\hline
40	&200	&52.36	&43.17	&36.42	&36.25	&51.28	&35.41	&41.42	&41.42	&45.40	&78.12		&31.06	&{\bf 30.94}\\
	&	&(1.52)	&(0.10)	&(0.36)	&(0.32)	&(1.44)	&(0.27)	&(0.01)	&(0.01)	&(0.41)	&(1.29)		&(0.05)	&(0.17)\\
	&500	&35.90	&40.52	&34.44	&34.32	&35.18	&33.56	&38.55	&38.55	&30.02	&67.15		&28.69	&{\bf 28.68}\\
	&	&(0.12)	&(0.03)	&(0.08)	&(0.05)	&(0.13)	&(0.11)	&(0.16)	&(0.17)	&(0.10)	&(2.50)		&(0.13)	&(0.09)\\
	&1000	&33.94	&35.19	&33.64	&33.58	&33.39	&32.87	&38.18	&38.18	&29.62	&31.75		&{\bf 27.95}	&28.00\\
	&	&(0.31)	&(0.25)	&(0.15)	&(0.13)	&(0.29)	&(0.12)	&(0.19)	&(0.19)	&(0.23)	&(1.28)		&(0.18)	&(0.13)
\end{tabular}
\caption{Simulation: The averaged LPDS values of the 12 density estimators for various dimension $d$ and number of observations $n$.
The numbers in brackets are standard deviations over the replications.
In each case, the minimum LPDS is in bold.}\label{T:simulation_LPDS}
\end{table}

\begin{table}[h]
\renewcommand{\arraystretch}{0.9} \centering \fontsize{8pt}{12pt}\selectfont
\centering
\vskip1mm
\begin{tabular}{cc|cccccccccccc}
$d$	&$n$	&MN	&M$t$	&MFA	&M$t$FA	&MAMN	&MAMFA	&NC	&$t$C	&CT-MN	&CT-M$t$	&CT-MFA	& CT-M$t$FA\\
\hline
5	&200	&0.03	&0.18	&1.39	&6	&0.16	&1.53	&0.01	&0.14	&17	&42		&102	&140\\
	&500	&0.09	&0.29	&12	&20	&0.21	&12	&0.04	&0.34	&40	&214		&456	&429\\
\hline
10	&200	&0.02	&0.29	&3.30	&8	&0.25	&3.53	&0.02	&0.20	&43	&145		&83	&251\\

	&500	&0.06	&0.35	&30	&43	&0.31	&30.8	&0.02	&0.60	&81	&381		&604	&967\\
\hline
40	&200	&0.03	&0.47	&8.46	&13	&0.65	&9.08	&0.02	&0.55	&43	&65		&103	&278\\

	&500	&0.12	&1.60	&51	&61	&1.06	&52	&0.03	&2.05	&184	&424		&421	&1148\\
	&1000	&0.24	&5.14	&91	&112	&2.07	&98	&0.08	&3.81	&201	&724		&672	&1634
\end{tabular}
\caption{Simulation: The CPU times (in seconds) of the 12 density estimators averaged over replications.
}\label{T:simulation_CPU}
\end{table}

%---------------------------------------------%
\subsection{Iris data}
\begin{figure}[h]
\centering
\includegraphics[width=1\textwidth,height =.5\textheight]{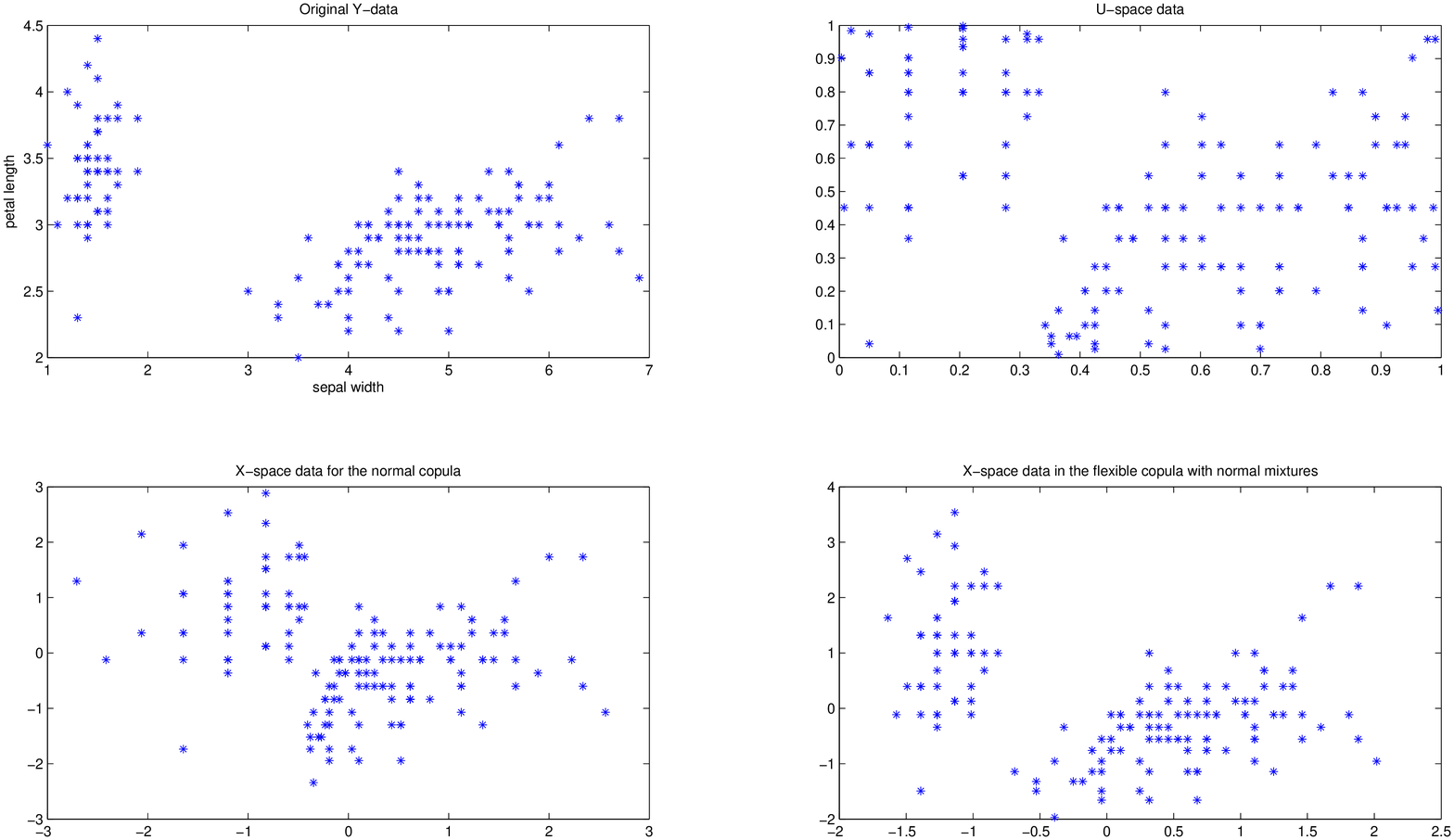}
\caption{Iris data: The top row shows the original $\v Y$-data of sepal width and petal length
and the $\v U$-space data. The lower-left panel shows the transformed $\v X$-space data
when the normal copula is used.
The last panel shows the $\v X$-space data (after the iterative scheme stops)
when the CT-MN model is used.}
\label{F:Iris}
\end{figure}

%---------------------------------------------%
This data set \citep{Fisher:1936} consists of observations of the lengths and widths of the sepals and petals of 150 Iris plants.
We are interested in estimating the joint density of these four variables.
For visualization purposes, we first consider the density estimation problem
in 2 dimensions, and estimate the joint density of the sepal width and the petal length.
The first row in Figure \ref{F:Iris} shows the original $\v Y$-data
and the $\v U$-space data, respectively.
We use the univariate mixture of $t$ model to estimate the marginals: a univariate $t$ mixture with two components is selected for the sepal width
and an univariate $t$ model is selected for the petal length.
The lower-left panel in Figure \ref{F:Iris} shows the transformed $\v X$-space data via $x_{ij}=\Phi^{-1}(u_{ij})$
when the normal copula is used.
If a normal copula is used then it is necessary to fit a bivariate normal to this data.
Clearly, it is unreasonable to do so.
The last panel shows the $\v X$-space data (after the iterative scheme stops) when we use
the CT-MN model.
A multivariate mixture of two normals is selected by the iterative scheme to estimate the dependence structure.
This mixture model seems to fit this data set well, visually showing
that the CT-MN model  captures the joint dependence structure in the data
better than the normal copula model.
Indeed, the 10-fold cross-validation LPDS values of
CT-MN and NC are $1.35$ and $1.70$, respectively.

\begin{table}[h]
\renewcommand{\arraystretch}{0.9} \centering \fontsize{10pt}{15pt}\selectfont
\centering
\vskip2mm
\begin{tabular}{c|cccccc}
\hline\hline
Estimators	&MN		&M$t$	  &MFA		&M$t$FA		&MAMN		&MAMFA\\
LPDS		&$1.74$	&$1.70$  &$2.73$	&$2.25$	&$1.71$	&$1.97$	\\
\hline
Estimators	&NC	  &$t$C		&CT-MN		&CT-M$t$	&CT-MFA		&CT-M$t$FA \\
LPDS		&$2.52$  &$2.52$	&$1.71$	&$\bf 1.67$	&$1.96$	&$2.01$	\\
\hline\hline
\end{tabular}
\caption{Iris data: 10-fold cross validation LPDS values for various estimators.
The minimum LPDS is in bold.}
\label{T:Iris}
\end{table}

We now consider estimating the joint density of all four variables,
and demonstrate the performance of various estimators using the LPDS criterion.
The best estimator for the first marginal is the implied mixture of normals,
and for the last three marginals the directly-estimated mixtures of $t$.
Table \ref{T:Iris} summarizes the 10-fold cross-validation LPDS values of these estimators.
We draw the following conclusions.
1) CT-M$t$ performs the best.
2) The copula-type estimators outperform the normal and $t$ copula estimators.
3) Dimension reduction via the factor analyzers models does not help,
probably because of the small dimension.
The improvement of the mixture-based copula-type estimators over the mixture estimators
shows that it is important to estimate the marginals separately.
The improvement of the copula-type estimators over the normal and $t$ copula estimators
shows that it is important to have flexibility in estimating the joint dependence.

%---------------------------------------------%
\subsection{Plasmodium gene expression data}
%---------------------------------------------%
Malaria is an infectious disease caused by the parasitic protozoan genus plasmodium.
This data set consists of the relative expression level of parasite genes taken at several time
points of the life cycle of parasites.
The original data set consisting of the expression level of 4221 genes taken at 46 time
points is further processed by \cite{Jasra:2007} using K-means clustering and principal component
analysis to reduce the number of observations from 4221 to 1000 and the number of variables
from 46 to 6. We use the processed data to demonstrate our proposed estimators.

The best estimators for the first three marginals are kernel densities
and for the last three are a mixture of $t$, a kernel density and a mixture of normals, respectively.
Table \ref{T:gene} summarizes the 5-fold cross validated LPDS values.
Typically we have the same conclusions as in the previous example:
1) the CT-M$t$ outperforms the others;
2) the copula-type estimators work better than the parametric copulas; and
3) dimension reduction does not help in this low-dimensional example.

\begin{table}[h]
\renewcommand{\arraystretch}{0.9} \centering \fontsize{10pt}{15pt}\selectfont
\centering
\vskip2mm
\begin{tabular}{c|cccccc}
\hline\hline
Estimators	&MN		&M$t$	  &MFA		&M$t$FA		&MAMN		&MAMFA\\
LPDS		&$10.76$	&$10.71$  &$11.65$	&$11.11$	&$10.73$	&$11.49$\\
\hline
Estimators	&NC	  &$t$C		&CT-MN		&CT-M$t$	&CT-MFA		&CT-M$t$FA \\
LPDS		&$11.84$  &$11.74$	&$10.73$	&${\bf 10.46}$	&$11.20$	&$10.84$\\
\hline\hline
\end{tabular}
\caption{Gene expression data: 5-fold cross validation LPDS values for various estimators.
The minimum LPDS is in bold.}
\label{T:gene}
\end{table}

%---------------------------------------------%
\subsection{Wine data set}
%---------------------------------------------%
This  data set consists of 13 chemical constituents found in 178 samples of wines in
a region of Italy. The data set and detailed information on it is available at the UC Irvine Machine
Learning Repository \texttt{http://archive.ics.uci.edu/ml/datasets/Wine}. The small number
of observations relative to the number of variables in this data set  shows the usefulness
of dimension reduction via the factor representation.
The best models for the marginals vary between a kernel density and a directly-estimated mixture of $t$ (details not shown).
Table \ref{T:wine} summarizes the multivariate model fitting results.
The best estimator is the CT-MFA.
In general,  dimension reduction  improve the performance,
e.g. the CT-MFA is better than the CT-MN, the CT-M$t$FA is better than the CT-M$t$.
The NC and $t$C work almost as well as the CT-MN and CT-M$t$ respectively.
This is because
the CT-MN and CT-M$t$ estimators  reduce to the NC and $t$C estimators respectively
when the joint dependence does not have a mixture structure.
\begin{table}[h]
\renewcommand{\arraystretch}{0.9} \centering \fontsize{10pt}{15pt}\selectfont
\centering
\vskip2mm
\begin{tabular}{c|cccccc}
\hline\hline
Estimators	&MN		&M$t$	  &MFA		&M$t$FA		&MAMN		&MAMFA\\
LPDS		&$19.67$	&$20.60$ &$20.53$	&$19.51$	&$19.24$	&$19.75$\\
\hline
Estimators	&NC	  &$t$C		&CT-MN		&CT-M$t$	&CT-MFA		&CT-M$t$FA \\
LPDS		&$19.22$  &$19.03$	&$19.22$	&$19.05$	&$\bf 18.96$	&$19.03$\\
\hline\hline
\end{tabular}
\caption{Wine data: 5-fold cross validation LPDS values for various estimators.
The minimum LPDS is in bold.}
\label{T:wine}
\end{table}

%======================================================================%
\section{Conclusion}\label{Sec:conclusion}
%======================================================================%
The article introduces copula-type estimators for flexible multivariate density estimation which
we believe improve on current popular copula estimators.
The new estimators allow the marginal densities to be modeled separately from the joint dependence,
as in all copula estimators, but have the ability to model complex joint dependence structures.
In particular, the joint dependence in the copula-type estimators that we propose
is modeled by mixture models.
The mixtures are fitted by Variational Bayes algorithms which automatically incorporate the model selection problem.
An iterative scheme is proposed for estimating copula-type estimators
and its usefulness is demonstrated through examples.

A practical issue is determining when a mixture-based copula-type estimator is needed for a given data set.
As can be seen from the examples, a mixture-based copula-type estimator works well
when the underlying joint dependence has a mixture structure.
Such an estimator can be obtained by the Variational Bayes fitting algorithm in our paper,
i.e. if the multivariate mixture $\widehat G$ estimated by the iterative scheme
has more than one component then it is likely that the underlying joint dependence has a mixture structure.
In our experience, if the underlying joint dependence does not have a mixture structure,
then the estimated multivariate mixture will have only one component
and the resulting copula-type estimator will be very similar to the corresponding normal or $t$ copula estimator.

In practice, it is necessary to select an estimator among the four mixture-based copula-type estimators proposed in the article.
In our experience, the CT-M$t$ often works well in small dimensions and the CT-M$t$FA is the best in high dimensions.
However, we suggest fitting all four estimators to the data
and then selecting the best estimator using some criterion such as the log predictive density score.

An alternative approach is to use marginally adjusted estimators \citep{Giordani:2012}
which try to improve on standard multivariate estimators such as a mixture of
multivariate normals, by modifying such estimators to take account of the best fitting marginal densities.
We believe that the copula-type and the marginal adaptation approaches complement each other,
in the sense that marginal adaptation attempts to correct deficiencies in standard
multivariate estimators and copula-type estimation attempts to make the popular copula models more flexible.
The practitioner may use both
approaches and choose the best performing one, by some criterion such as the log predictive score.

We note, however, that if we wish to incorporate dependence on the covariates in the marginals, then it is easier
to do so using the copula type estimators than the marginally adjusted estimators because of the need to estimate the
normalizing constants in the marginally adjusted estimators.

%======================================================================%
\section*{Acknowledgment}
%======================================================================%
The authors would like to thank the referees for insightful comments which helped to improve the
presentation and content of the paper.
The research of Minh-Ngoc Tran, Xiuyan Mun and Robert Kohn
was partially supported by Australian Research Council grant DP0667069.
We thank Professor Ajay Jasra for the genome data.

%======================================================================%
\section*{Appendix A}
%======================================================================%
\begin{proof}[Proof of Proposition 1]
Without loss of generality, assume that $d=2$.
By construction, $H_i(X_i)=F_i(Y_i)$, $i=1,2$ and $\v X=(X_1,X_2)'\sim \widehat G(x_1,x_2)$.
The distribution $\widehat F(\v y|H,\widehat G)$ of $\v Y=(Y_1,Y_2)'$ is
\bean
\widehat F(\v y|H,\widehat G)&=&\P(Y_1\leq y_1,Y_2\leq y_2)\\
&=&\P\Big(X_1\leq H_1^{-1}(F_1(y_1)),X_1\leq H_2^{-1}(F_2(y_2))\Big)\\
&=&\widehat G\big(H_1^{-1}(F_1(y_1)),H_2^{-1}(F_2(y_2))\big)\\
&=&\widehat C(F_1(y_1),F_2(y_2)|H,\widehat G).
\eean
Taking derivatives with respect to $y_1$ and $y_2$,
we obtain the density function of $\v Y$
\beqn
\widehat f(\v y|H,\widehat G)=\widehat g(x_1,x_2)\prod_{i=1}^2\frac{f_i(y_i)}{h_i(x_i)}
\eeqn
with $H_i(x_i)=F_i(y_i)$.

Now noting that $h_i(x_i)dx_i=f_i(y_i)dy_i$, the density of the first marginal $Y_1$ is
\beqn
\widehat f_1(y_1|H,\widehat G)=\int\widehat f(\v y|H,\widehat G)dy_2=\int\widehat g(x_1,x_2)\frac{f_1(y_1)}{h_1(x_1)}dx_2=\widehat g_1(x_1)\frac{f_1(y_1)}{h_1(x_1)}
\eeqn
\end{proof}

\begin{proof}[Proof of Proposition 2]
By construction, the data $\D_X^{H_{1:d}}$ are realizations of a random vector $\v X=(X_1,...,X_d)'$
obtained by the transformation $X_i=H_i^{-1}(U_i)$ with $U_i$ uniformly distributed on $[0,1]$, $i=1,...,d$.
Therefore $h_i(x_i)$ are the marginal pdf's of $\v X$.
Denote by $h(\v x)$ be the joint pdf of $\v X$, we have that
\beqn
h_i(x_i) = \int h(\v x)d\v x_{-i},\;\;\text{with}\;\;\v x_{-i}=(x_1,...,x_{i-1},x_{i+1},...,x_d)'.
\eeqn
Noting that $h_i(x_i)dx_i=f_i(y_i)dy_i$, by the second result in Proposition 1,
\begin{align*}
d_\text{TV}(\widehat f_i,f_i)&=\frac12\int\vert \widehat f_i(y_i)-f_i(y_i)\vert dy_i\\
&\leq \frac12\int|\widehat g_i(x_i)-h_i(x_i)|dx_i\\
& = \frac12 \int_{x_i}  \mid  \left ( \int_{{\v x}_{-i} } \left ( {\widehat g} (\v x) - h(\v x) \right ) d {\v x}_{-i}\right )  \mid  dx_i \\
&\leq \frac12\int|\widehat g(\v x)-h(\v x)|d\v x\to 0,
\end{align*}
when the sample size inncreases,
because the fitting method is reliable.
\end{proof}

%======================================================================%
\section*{Appendix B: Variational Bayes algorithms for fitting mixture models}
%======================================================================%
Using Variational Bayes for fitting mixture models has proven useful and efficient.
See, e.g., \cite{Ormerod:2009} for an introduction to Variational Bayes.
\cite{Giordani:2012} develop efficient Variational Bayes algorithms for fitting a multivariate mixture of normals
and a mixture of factor analyzers in which the number of components and the number of factors in each component are automatically selected.
We present here Variational Bayes algorithms for fitting a multivariate mixture of $t$
and a mixture of $t$-factor analyzers,
in which the model selection problem is also automatically incorporated.

%---------------------------------------%
\subsection*{Fitting a mixture of $t$}
%---------------------------------------%
The density of the mixture of $t$ model is of the form
\beq\label{tmixture}
p(\v x) = \sum_{k=1}^K\pi_kt_d(\v x;\v\mu_k,V_k,\nu_k),
\eeq
where $t_d(\v x;\v\mu,V,\nu)$ denotes the density of a $d$-variate $t$ distribution
with location $\v\mu$, scale matrix $V$ and degrees of freedom $\nu$.
The mean of this $t$ distribution is $\v\mu$ if $\nu>1$, and its variance matrix is $(\nu/(\nu-2))V$ if $\nu>2$.
The key to our Variational Bayes fitting approach is the expression of $t$ distributions as scale mixtures of normals \citep{Andrews:1974}.
The distribution of $X\sim t_d(\v x;\v\mu,V,\nu)$ can be expressed hierarchically as
\beqn
X|w\sim N_d(\v\mu,V/w)\;\;\text{with}\;\;w\sim\mathcal G\left (\frac{\nu}{2},\frac{\nu}{2}\right ).
\eeqn
Using this result, the model \eqref{tmixture} can be written as
\bean
\v x_i|\d_i=j,w_{ij}&\sim&N_d(\v\mu_j,(w_{ij}T_j)^{-1})\\
p(\d_i=j)&=&\pi_j\\
w_{ij}&\sim&\mathcal G\left (\frac{\nu_j}{2},\frac{\nu_j}{2}\right ),\;\; i=1,...,n,\;\; j=1,...,K
\eean
with $\d_i$ and $w_{ij}$ latent variables.
Here $T_j=V_j^{-1}$. For now we consider the degrees of freedom $\v\nu=(\nu_1,...,\nu_K)$ as fixed hyperparameters.
This will be relaxed below.
The model parameters are $\v\t=(\v\pi,\v w,\v\d,T,\v\mu)$.
We consider the following decomposition
\beq\label{E:decomposition}
p(\v\t)=p(\v\pi)p(\v\d|\v\pi)p(\v w)p(T)p(\v\mu|T)
\eeq
with the conjugate priors
\bean
p(\v\pi)&\sim&\text{Dirichlet}(\v\a^0)\\
p(\v w)&=&\prod_{i=1}^n\sum_{j=1}^K1_{\d_i=j}p(w_{ij})\\
p(\v\d|\v\pi)&\sim&\prod_{i=1}^n\sum_{j=1}^K1_{\d_i=j}\pi_j\\
T_j&\sim&\text{Wishart}(\tau_j^0,{\Sig_j^0}^{-1})\\
\v\mu_j|T_j&\sim&N_d(0,(\k^0_jT_j)^{-1}),
\eean
where $\v\a^0$, $\tau_j^0$, $\Sig_j^0$ and $\k_j^0$ are hyperparamters.
Note that at the moment the degrees of freedom $\nu_j$ are also considered as hyperparameters.
From the decomposition \eqref{E:decomposition} \citep[cf.][]{Ormerod:2009}, the optimal Variational Bayes posteriors are
\bean
q_{ij}=q(\d_i=j)&\propto&\exp\Big([\log\pi_j]+(\frac{\nu_j}{2}+\frac d2-1)[\log w_{ij}]\\
&&\phantom{ccccccc}-(\frac{\nu_j}{2}+\frac{z_{ij}}{2})[w_{ij}]+\frac12[\log|T_j|]+\frac{\nu_j}{2}\log\frac{\nu_j}{2}-\log\G(\frac{\nu_j}{2})\Big)\\
q(\v\pi)&\sim&\text{Dirichlet}(\v\a)\;\;\text{with}\;\;\a_j=\a_j^0+\sum_iq_{ij}\\
q(w_{ij})&\sim&\mathcal G\left(\frac{\nu_j}{2}+\frac d2,\frac{\nu_j}{2}+\frac{z_{ij}}{2}\right)\\
q(\v\mu_j|T_j)&\sim&N_d(\v\mu_j^q,(\k_jT_j)^{-1})\\
&&\k_j=\k_j^0+\sum_iq_{ij}[w_{ij}],\;\;\v\mu_j^q=\frac{1}{\k_j}\sum_iq_{ij}[w_{ij}]\v x_i\\
q(T_j)&\sim&\text{Wishart}(\tau_j,\Sig_j^{-1}),\;\;\;\tau_j=\tau_j^0+1+\sum_iq_{ij}\\
&&\Sig_j=\Sig_j^0+\k_j^0\v\mu_j^q(\v\mu_j^q)'+\sum_iq_{ij}[w_{ij}](\v x_i-\v\mu_j^q)(\v x_i-\v\mu_j^q)'
\eean
where $[.]$ denotes expectation with respect to the Variational Bayes posterior $q$, i.e., $[.]:=E_q(.)$.
In the above
\bean
z_{ij}=[(\v x_i-\v\mu_j)'T_j(\v x_i-\v\mu_j)]=\tau_j(\v x_i-\v\mu_j^q)'\Sig_j^{-1}(\v x_i-\v\mu_j^q)+\frac{d}{\k_j}
\eean
and $[\log\pi_j]=\Psi(\a_j)-\Psi(\sum_j\a_j)$, $[\log w_{ij}]=\Psi(\frac{\nu_j}{2}+\frac d2)-\log (\frac{\nu_j}{2}+\frac{z_{ij}}{2})$,
$[w_{ij}]=(\frac{\nu_j}{2}+\frac d2)/(\frac{\nu_j}{2}+\frac{z_{ij}}{2})$ and
$[\log|T_j|]=\sum_{h=1}^d\Psi(\frac12(\tau_j+1-h))+d\log2-\log|\Sig_j|$.
Let $L_1(\v\nu)$ be the lower bound on $\log p(\v x|\v\nu)$.

Estimating the degrees of freedom is challenging in both Bayesian and frequentist approaches.
In our setting, the optimal Variational Bayes posterior of $\nu_j$ does not have any standard form.
We proceed as follow.
Let $p(\v\nu)$ be a prior on $\v\nu$. We use a point mass distribution for the Variational Bayes posterior of $\v\nu$, i.e., $q(\v\nu)=\d(\v\nu-\v\nu^q)$
with $\d(.)$ the Dirac delta distribution.
The lower bound on $\log p(\v x)$ is
\beq
\int\log\frac{p(\v\nu)p(\v x|\v\nu)}{q(\v\nu)}q(\v\nu)d\v\nu=\log p(\v\nu^q)+\log p(\v x|\v\nu^q).
\eeq
With $L_1(\v\nu^q)$ the lower bound on $\log p(\v x|\v\nu^q)$, the lower bound on $\log p(\v x)$ is
\beq\label{equ:finallb}
L=\log p(\v\nu^q)+L_1(\v\nu^q).
\eeq
This needs to be optimized with respect to $\v\nu^q$.
We will use the notation $\v\nu$ instead of $\v\nu^q$ in what follows.

It is well known in Bayesian fitting of $t$ distributions that an improper prior on the degrees of freedom
leads to an improper posterior,
while in frequentist fitting the MLE may not converge because of the non-regularity of the likelihood.
A truncated prior is commonly used.
We follow \cite{Lin:2004} and use the uniform prior on $(0,\l^0)$,
with some sufficiently large $\l^0$, say $\l^0=100$.
Then maximizing \eqref{equ:finallb} is equivalent to maximizing the following function in $\nu_j$
\beq
\sum_{i=1}^nq_{ij}\Big(\frac{\nu_j}{2}\log(\frac{\nu_j}{2})-(\frac{\nu_j}{2}+\frac d2)\log(\frac{\nu_j}{2}+\frac{z_{ij}}{2})+\log\G(\frac{\nu_j}{2}+\frac d2)-\log\G(\frac{\nu_j}{2})\Big)
\eeq
subject to $\nu_j\in[0,\l^0]$, $j=1,...,K$.
This is somewhat similar to the M-step update of the degrees of freedom in the EM algorithm of \cite{Peel:2000}.
However, \cite{Peel:2000} did not impose any constraint on $\nu_j$,
which may cause divergence of the solution.
For simplicity, we consider $\nu_j$ to be integer.

Given an initial number of components $K$, the Variational Bayes algorithm
sequentially updates the parameters $q_{ij}$, $\a_j$, $\kappa_j$, $\v\mu_j^q$, $\Sig_j$ and $\nu_j$
until some stopping rule is met.
Often, this iterative scheme stops
when the lower bound \eqref{equ:finallb} is not improved any further,
or when the updates are stable in the sense that
the difference of main parameters $\v\mu_j^q$ and $\Sig_j$ in two successive iterations is
smaller than a tolerance value.
We refer to this update procedure as the standard Variational Bayes algorithm.

To select $K$, we start with a reasonably large value of $K$
and remove redundant components on the basis of maximizing the lower bound as follows.
After the standard Variational Bayes procedure has converged, we try removing the components with smallest $\sum_i q_{ij}$
and actually remove these components if the final optimized lower bound is improved.
That is, unlike the existing algorithms
in which components with the posterior probabilities $\frac1n\sum_i q_{ij}$
smaller than a specific threshold value are eliminated  \citep{Corduneanu:2001,McGrory:2007},
we first rank components for elimination and eliminate plausible components
until the lower bound is not improved any further.
We found that our strategy quickly and efficiently eliminates redundant components,
while not requiring any specific threshold value which may be hard to determine.
We will refer to this algorithm for determining $K$ as the Elimination Variational Bayes (EVB).
It might be desirable to include split steps which split poorly-fitted components.
However, implementation of split steps is difficult in the $t$ mixture context
because it is not clear how to initialize new components optimally.

%--------------------------------------%
\subsection*{Fitting a mixture of $t$-factor analyzers}
%--------------------------------------%
The density of a mixture of $t$-factor analyzers is \eqref{tmixture}
with the scale matrices having factor representation $V_k=\Ld_k\Ld_k'+\Psi_k$.
This model is first considered in \cite{McLachlan:2007} who develop an EM algorithm for fitting.
Model selection in fitting this model consists of
selecting the number of components $K$ and the number of factors $\g_k$ in each component.
Therefore the number of models in the model space is huge,
which makes the model selection problem challenging
when using model selection criteria such as AIC or BIC because one needs to search over the whole model space.
To reduce the model space, \cite{McLachlan:2007} consider the same number of factors $\g_j\equiv \g$ for all components.
We relax this assumption here
and develop below a Variational Bayes algorithm for fitting the model
in which $K$ and $\g_k$ are automatically determined.
We believe this is the first algorithm in the literature for fitting the (full) mixture of $t$-factor analyzers model
which is able to do parameter estimation and model selection simultaneously and automatically.

The model can be written as
\bean
\v x_i|\d_i=j,\v z_{ij},w_{ij}&\sim&N_d(\v\mu_j+\Ld_j\v z_{ij},(w_{ij}\psi_j)^{-1}I)\\
\v z_{ij}&\sim&N_{k_j}(\v0,w_{ij}^{-1}I)\\
w_{ij}&\sim&\mathcal{G}(\frac{\nu_j}{2},\frac{\nu_j}{2})\\
p(\d_i=j)&=&\pi_j
\eean
with $\v z_{ij}$, $w_{ij}$, $\d_i$ latent variables.
Following \cite{McLachlan:2007} we assume $\Psi_j=\psi_j^{-1}I$,
which helps avoid spikes or near singularities in the likelihood.
We consider the following priors on the model parameters
\bean
p(\v\pi)\sim\text{Dirichlet}(\v\a^0),\;\;p(\v\mu_j)\sim1,\;\;p(\Ld_j|\tau_j)=\prod_{l=1}^{k_j}N_d(\v0,\tau_{jl}^{-1}I)
\eean
with $\tau_j=(\tau_{j1},...,\tau_{jk_j})$
and put gamma priors $\mathcal G(a,b)$ on $\tau_{jl}$ and $\psi_j$.
The form of the prior $p(\Ld_{j}|\tau_j)$ plays a key role in determining the local dimensions $k_j$:
a very small value of $\tau_{jl}^{-1}$ suggests that the factor $l$
of the component $j$ should be removed.
This approach is introduced in \cite{Ghahramani:2000} for mixtures of (normal) factor analyzers.

The Variational Bayes optimal posteriors for the parameters are as follows
\bean
q(\v z_{ij})&\sim&N_{k_j}(\v\mu_{x_{ij}},\Sig_{x_{ij}})\\
&&\Sig_{x_{ij}}=[w_{ij}]^{-1}(I+[\psi_j][\Ld_j'\Ld_j])^{-1},\;\;\v\mu_{x_{ij}}=\Sig_{x_{ij}}[w_{ij}][\psi_j][\Ld_j'](\v x_i-\v\mu_j^q)\\
q(w_{ij})&\sim&\mathcal G\left(\frac{\nu_j}{2}+a_{w_{ij}},\frac{\nu_j}{2}+b_{w_{ij}}\right)\\
&&a_{w_{ij}}=\frac{k_j}{2}+\frac d2,\;\;b_{w_{ij}}=\frac12[\psi_j]c_{ij}+\frac12\v\mu_{x_{ij}}'\v\mu_{x_{ij}}+\frac12\tr(\Sig_{x_{ij}})\\
q_{ij}&\propto&\exp\Big([\log\pi_j]+(\frac{\nu_j}{2}+\frac{k_j}{2}+\frac d2-1)[\log w_{ij}]+\frac d2[\log\psi_j]\\
&&\phantom{ccccccccccccc}-(\frac{\nu_j}{2}+\frac{k_j}{2}+\frac d2)-\frac{k_j}{2}\log(2\pi)+\frac{\nu_j}{2}\log(\frac{\nu_j}{2})-\log\G(\frac{\nu_j}{2})\Big)\\
q(\v\pi)&\sim&\text{Dirichlet}(\v\a),\;\;\a_j=\a_j^0+\sum_{i=1}^nq_{ij}\\
q(\v\mu_j)&\sim&N_d(\v\mu_{\mu_j},\sigma_{\mu_j}^2I),\\
&&\sigma_{\mu_j}^{2}=\left([\psi_j]\sum_{i=1}^nq_{ij}[w_{ij}]\right)^{-1},
\;\v\mu_{\mu_j}=\sigma_{\mu_j}^2[\psi_j]\sum_{i=1}^nq_{ij}[w_{ij}](\v x_i-[\Ld_j][\v z_{ij}])\\
q(\v\Ld_{jl})&\sim&N_d(\v\mu_{\Ld_{jl}},\s_{\Ld_{jl}}^2I),\\
&&\s_{\Ld_{jl}}^2=\left([\tau_{jl}]+[\psi_j]\sum_{i=1}^nq_{ij}[w_{ij}][x_{ij,l}^2]\right)^{-1},\\
&&\v\mu_{\Ld_{jl}}= \s_{\Ld_{jl}}^2[\psi_j]\sum_{i=1}^nq_{ij}[w_{ij}]\Big([x_{ij,l}](\v x_i-\v\mu_{\mu_j})-\sum_{s\not=l}\v\mu_{\Ld_{js}}[x_{ij,l}x_{ij,s}]\Big)\\
q(\tau_{jl})&\sim&\mathcal G(a_{\tau_{jl}},b_{\tau_{jl}})\\
&&a_{\tau_{jl}}=a+\frac d2,\;\;b_{\tau_{jl}}=b+\frac12\v\mu_{\Ld_{jl}}'\v\mu_{\Ld_{jl}}+\frac{d}{2}\s^2_{\Ld_{jl}}\\
q(\psi_j)&\sim&\mathcal G(a_{\psi_{j}},b_{\psi_{j}})\\
&&a_{\psi_{j}}=a+\frac d2\sum_{i=1}^nq_{ij},\;\;b_{\psi_{j}}=b+\frac12\sum_{i=1}^nq_{ij}[w_{ij}]c_{ij}\\
\eean
where
\beqn
c_{ij}=(\v x_i-\v\mu_{\mu_j})'(\v x_i-\v\mu_{\mu_j})-2(\v x_i-\v\mu_{\mu_j})'[\Ld_j]\v\mu_{x_{ij}}+\v\mu_{x_{ij}}'[\Ld_j'\Ld_j]\v\mu_{x_{ij}}+\frac d2\s_{\mu_j}^2+\tr(\Sigma_{x_{ij}}[\Ld_j'\Ld_j]).
\eeqn
The expectation terms are given by
\beqn
[\Ld_j'\Ld_j]=[\Ld_j]'[\Ld_j]+\frac d2\diag(\s_{\Ld_{j1}}^2,...,\s_{\Ld_{jk_j}}^2)
\eeqn
and
\beqn
[x_{ij,l}x_{ij,s}]=(\Sig_{x_{ij}})_{l,s}+\v\mu_{x_{ij}}^{(l)}\v\mu_{x_{ij}}^{(s)}.
\eeqn
Similar to the reasoning in the previous section,
the degrees of freedom $\nu_j$ are estimated by maximizing
\beqn
\sum_{i=1}^nq_{ij}\Big(\frac{\nu_j}{2}\log(\frac{\nu_j}{2})-(\frac{\nu_j}{2}+a_{w_{ij}})\log(\frac{\nu_j}{2}+b_{w_{ij}})+\log\G(\frac{\nu_j}{2}+a_{w_{ij}})-\log\G(\frac{\nu_j}{2})\Big)
\eeqn
subject to $\nu_j\in[0,\l^0]$, $j=1,...,K$.

Our standard Variational Bayes algorithm sequentially updates the parameters $\Sig_{x_{ij}}$, $\v\mu_{x_{ij}}$,
$a_{w_{ij}}$, $b_{w_{ij}}$, $q_{ij}$, $\a_j$, $\sigma^2_{\mu_j}$, $\v\mu_{\mu_j}$,
$\sigma^2_{\Ld_{jl}}$, $\v\mu_{\Ld_{jl}}$, $a_{\tau_{jl}}$, $b_{\tau_{jl}}$, $a_{\psi_j}$, $b_{\psi_j}$ and $\nu_j$
until the difference of main parameters $\v\mu_{\mu_j}$ and $\v\mu_{\Ld_{jl}}$ in two successive iterations is
smaller than a tolerance value.
Other stopping rules can be used as well.

We now present our strategy for determining the local dimensions $k_j$.
We remove the factor $l$ of the component $j$ if the posterior mean of $\tau_{jl}^{-1}$
is smaller than a threshold $\epsilon$.
Note that the mean of $\tau_{jl}^{-1}$ is $b_{\tau_{jl}}/(a_{\tau_{jl}}-1)$.
Because the unit of these means depends on that of the data $\v x$,
we found it necessary to standardize the data such that the columns of $\v x$ have standard deviations of 1;
this makes the analysis more stable and facilitates the choice of $\epsilon$.
After fitting, it is straightforward to write the resulting density back in the original units.
From our experience, $\epsilon=10^{-3}$ is a good choice.
To select $K$, we follow the same elimination Variational Bayes strategy as in the previous section.

In summary, our strategy for model selection in fitting the M$t$FA model is as follows.
\begin{itemize}
\item Step 1: Start with a reasonably large value of $K$ and with the initial number of factors
$k_j=[\frac12(2d+1-\sqrt{8d+1})]$ - the largest value allowed for the number of
factors in factor analysis.
\item Step 2: After the standard Variational Bayes procedure has converged, remove factors with $b_{\tau_{jl}}/(a_{\tau_{jl}}-1)<\epsilon$.
\item Step 3: Remove redundant components via the EVB algorithm.
\item Step 4: Repeat steps 2 and 3 until the lower bound is not improved any further.
\end{itemize}

\bibliographystyle{apalike}
\bibliography{t_mixtures_ref}

\end{document}